\documentclass[prd,aps,superscriptaddress,floatfix,nofootinbib,eqsecnum,twocolumn]{revtex4-2}

\pdfoutput=1

\usepackage{amsfonts}
\usepackage{amsmath}
\usepackage{amssymb}
\usepackage{bm}
\usepackage{dcolumn}
\usepackage{graphicx}   
\usepackage{latexsym}
\usepackage{rotating}   
\usepackage{hyperref}
\usepackage{subfigure}
\usepackage{color}
\usepackage{changes}
\usepackage{verbatim}
\usepackage{comment}
\usepackage{empheq}
\usepackage{physics}
\usepackage{amsmath,latexsym}
\usepackage{float}

\begin{document}

\title{Constraining a de Broglie-Bohm quantum bounce cosmology with Planck data}

\author{Micol Benetti}
\affiliation{Scuola Superiore Meridionale, Via Mezzocannone 4, I-80138, Naples, Italy.}
\affiliation{INFN Sez. di Napoli, Compl. Univ. di Monte S. Angelo, Edificio G, Via	Cinthia, I-80126, Naples, Italy.}
\email{m.benetti@ssmeridionale.it}

\author{Rudnei O. Ramos}
\email{rudnei@uerj.br}
\affiliation{Departamento de F\'{\i}sica Te\'orica, Universidade do Estado do Rio de Janeiro, 20550-013 Rio de Janeiro, RJ, Brazil}

\author{Renato Silva}
\email{renatobqa@gmail.com}
\affiliation{Departamento de F\'{\i}sica Te\'orica, Universidade do Estado do Rio de Janeiro, 20550-013 Rio de Janeiro, RJ, Brazil}

\author{Gustavo S. Vicente}
\email{gustavo@fat.uerj.br}
\affiliation{Faculdade de Tecnologia, Universidade do Estado do Rio de Janeiro, 27537-000 Resende, RJ, Brazil}

\begin{abstract}

This work investigates a singularity-free early Universe within the paradigm of quantum cosmology. We develop a bouncing model where the singularity is resolved via the de Broglie-Bohm interpretation of quantum mechanics, which provides a deterministic trajectory for the scale factor through a quantum bounce. The primordial power spectrum for scalar perturbations is derived, incorporating a characteristic modulation (distortion function) imprinted by the non-standard quantum gravitational dynamics near the bounce. We confront this model with the Planck 2018 cosmic microwave background data, establishing its strong compatibility with observations. Our analysis places a stringent upper bound on the fundamental scale of the bounce $k_B$, constraining the parameter space of such quantum cosmological scenarios. {}Furthermore, the model's specific scale-dependent anti-correlation between the spectral index and amplitude of perturbations offers a potential mechanism for mitigating the $H_0$-$\sigma_8$ tension, presenting a testable signature for future cosmological surveys.

\end{abstract}

\maketitle 



\section{Introduction}
\label{intro}

Modern cosmology is fundamentally grounded in the hot Big Bang model, which describes a dynamic, expanding Universe that emerged from an ultra-hot and dense primordial state. This framework enjoys significant success, providing a robust explanation for fundamental observations, including the cosmic microwave background (CMB) radiation, the abundance of light elements from Big Bang nucleosynthesis (BBN), and the large-scale structure (LSS) of the Universe. Nevertheless, this model is inherently incomplete, as it extrapolates back to an initial spacetime singularity --- a point where the predictions of general relativity break down and energy densities diverge. This singularity represents not merely a mathematical artifact but the fundamental boundary of the theory's validity, necessitating a paradigm shift to a quantum theory of gravity for a complete description of the origin of the Universe.

A cornerstone of modern cosmology is the presence of a primordial epoch of accelerated expansion, known as cosmic inflation. This paradigm elegantly resolves key fine-tuning problems of the standard Big Bang model --- namely the horizon, flatness, and magnetic monopole problems~\cite{Guth:1980zm,Linde:1983gd,Albrecht:1982wi}. {}Furthermore, it provides a powerful mechanism for generating the spectrum of primordial quantum fluctuations that seed the large-scale structure observed in the Universe today. Despite its remarkable successes, the standard inflationary framework is typically constructed on a classical spacetime geometry, implicitly assuming its validity immediately after the Planck epoch. Consequently, it does not address the singularity problem and remains divorced from a first-principles quantum gravitational description of the Universe's very origin.

Bouncing models~\cite{Khoury:2001wf,Khoury:2003rt,Biswas:2006bs,Peter:2008qz,Novello:2008ra,Cai:2011zx,Cai:2012va,Battefeld:2014uga,Schander:2015eja,Brandenberger:2016vhg,Ilyas:2020qja,Pinto-Neto:2021gcl,Delgado:2021mxu,Ageeva:2021yik,Ageeva:2024knc} provide a compelling alternative to cosmic inflation by resolving the initial singularity problem while preserving its observational successes. Quantum cosmology emerges as a natural framework for such models, as the classical description of spacetime breaks down at the high energies where the bounce occurs, necessitating a quantum gravitational treatment. In this picture, the initial singularity is replaced by a non-singular quantum bounce, forming a bridge between a preceding contracting phase and our observed expanding universe.

{}Furthermore, while bouncing models can, in principle, be considered alternatives to inflation, they are not mutually exclusive. A particularly compelling and widely studied scenario is a hybrid approach in which a quantum bounce is followed by a period of inflation. This synthesis addresses the initial singularity problem through the bounce while simultaneously leveraging inflation's proven successes, such as generating the large-scale structure of the universe. Among the most rigorously developed approaches in this vein is Loop Quantum Cosmology (LQC), which applies loop quantum gravity techniques to homogeneous spacetimes~\cite{Ashtekar:2006uz,Bojowald:2005epg,Ashtekar:2021kfp}. It is important to note that a primary application of LQC is not merely to provide a standalone bounce, but to study the conditions for and dynamics of a subsequent inflationary epoch. Consequently, LQC is very commonly investigated within the context of this "bounce-plus-inflation" paradigm, where the quantum-gravitational bounce provides natural initial conditions for the inflaton field~\cite{Ashtekar:2011rm,Bonga:2015xna,Bhardwaj:2018omt}.
 
In LQC, the repulsive character of quantum geometry at near-Planckian densities naturally generates a bounce, thus avoiding the singularity. Crucially, this quantum gravitational transition is not merely a theoretical artifact, but can imprint specific signatures on primordial perturbations. As demonstrated, for example, in Refs.~\cite{Agullo:2013ai,Wilson-Ewing:2016yan,Ashtekar:2016wpi,MenaMarugan:2024zcv,Montese:2024ypz}, these may include a suppression of power spectrum at large angular scales in the CMB and characteristic modifications to the primordial gravitational wave spectrum, rendering the paradigm possibly testable through future CMB observations.

{Complementary to these geometrically motivated approaches like LQC, another framework seeks to resolve the singularity through the direct quantization of the Universe's wave function, governed by the Wheeler-DeWitt (WDW) equation. 
The fundamental interpretative challenges of the WDW equation --- its timeless nature and the problem of quantum measurement for a closed Universe --- motivate the adoption of alternative quantum interpretations. 
The de Broglie-Bohm (dBB) pilot-wave theory~\cite{Bohm:1951xw,Bohm:1951xx} is particularly well suited for this task. 
The dBB formulation of quantum cosmology provides a deterministic description of quantum evolution via well-defined trajectories, guided by the phase of the universal wave function, where the Universe is described by a single, well-defined geometric configuration guided by the wave function through the quantum potential.
Even when the wave function is highly nonclassical, the background evolution follows a unique Bohmian trajectory, ensuring a definite spacetime geometry. 
As a result, the background geometry is never in a superposition, which guarantees the consistency of treating perturbations as quantum fields propagating on a definite spacetime.
This property allows cosmological perturbations to be consistently quantized and evolved on an effective semiclassical background, in close analogy with other semiclassical treatments in quantum cosmology~\cite{Halliwell:1984eu}. Crucially, the dBB interpretation furnishes an objective, observer-independent description of quantum dynamics without invoking the collapse postulate, thereby offering a coherent ontological framework for quantum cosmology and a potential mechanism for a non-singular bounce~\cite{Pinto-Neto:2021gcl,Pinto-Neto:2021jko}.}

Building upon this dBB framework, in Ref.~\cite{Vicente:2023hba} a concrete cosmological model featuring a stiff-matter dominated quantum bounce that transitions smoothly into a standard inflationary phase was developed. The model, which considers a flat {}Friedmann-Lema\^itre-Robertson-Walker (FLRW) Universe with a scalar field, yields complete analytical solutions spanning the entire history from quantum contraction through the bounce to the end of inflation. This analytical tractability provides a powerful tool for investigating how fundamental quantum parameters of the pre-bounce wave function --- such as the characteristic bounce density and wave packet dispersion --- determine the number of inflationary e-folds and thereby set the initial conditions for the formation of large-scale structure.

Crucially, the analysis in~\cite{Vicente:2023hba} identified specific regions of this parameter space that produce predictions consistent with Planck data, demonstrating that a unified bounce-plus-inflation scenario is not only theoretically viable, but also empirically constrained. Consequently, such hybrid models represent a more complete conceptual framework for the early Universe. By inherently avoiding the initial singularity, they provide a direct link between Planck-scale quantum gravity and late-time cosmological observations, providing a possible way for a new generation of observational tests on the origin of the Universe.

In this work, we extend the analysis of quantum cosmological bounce models within the dBB framework to the dynamics of primordial perturbations. Our primary objective is to compute the precise evolution of quantum fluctuations across the bounce and derive their resulting observational signatures. To this end, we encode the modifications imprinted on the primordial power spectrum by the quantum-gravitational bounce dynamics into a novel, phenomenological distortion function that is derived in the dBB framework. This function is integrated into a modified version of the Cosmic Anisotropy Solving System Boltzmann code \texttt{CAMB}~\cite{Lewis:1999bs}, enabling the computation of the corresponding CMB anisotropy spectra.

We then performed a comprehensive statistical analysis of the model using the Cobaya framework~\cite{Torrado:2020dgo}, using its Markov Chain Monte Carlo (MCMC) sampler to conduct a complete parameter inference against Planck 2018 data. This methodology allows us to efficiently explore the high-dimensional parameter space, which includes both standard cosmological and novel bounce parameters. Our results place stringent constraints on the key parameters characterizing the bounce model, specifically the energy scale at which the bounce occurs, demonstrating the power of cosmological data to probe physics in the quantum-gravitational regime.

This paper is organized as follows. In Sec.~\ref{section2}, we introduce the background cosmological model, which is the standard flat FLRW Universe filled with a scalar field, and we also briefly describe the quantum model in the context of the dBB interpretation. 
In Sec.~\ref{section3}, we give the calculation of the primordial scalar perturbations in the model studied here.
In Sec.~\ref{section4}, we provide a statistical analysis of the perturbations of the model, which allows us to constrain the quantum bounce parameters. Our conclusions and future perspectives are given in Sec.~\ref{section5}.

\section{Background Cosmological Model}
\label{section2}

We consider a flat FLRW Universe with line element
\begin{equation}
ds^2 = -dt^2 + a^2(t)d{\bf x}^2 .
\end{equation}
The classical and quantum background dynamics follow closely the treatment of Ref.~\cite{Vicente:2023hba}. 
{}For completeness, we summarize only the elements required for the present analysis and point to Ref.~\cite{Vicente:2023hba} for full derivations.

\subsection{Classical Model}

We take the matter content to be a canonical scalar field with potential $V(\phi)$, described by the action
\begin{equation}
\mathcal{S} = \int d^{4}x \sqrt{-g} \left[\frac{R}{2\kappa}
- \frac12 g^{\mu\nu}\partial_\mu\phi\partial_\nu\phi - V(\phi)\right],
\end{equation}
with $\kappa=\sqrt{8\pi}/m_{\rm Pl}$,  $m_{\rm Pl}$ is the
Planck mass ($m_{\rm Pl}\simeq 1.22 \times 10^{19}\,$GeV),
$g_{\mu\nu}$ is the metric tensor,  $g$ is the determinant of the
metric tensor and $R$ is the Ricci scalar.   
The Hamiltonian density for a homogeneous field in a flat FLRW geometry is
\begin{equation}
\mathcal{H}_\phi = -\kappa^2\frac{P_a^2}{12a}+
\frac{P_\phi^2}{2a^3}+a^3V(\phi),
\end{equation}
as also written in Ref.~\cite{Vicente:2023hba}. In the above equation, $P_a$ and $P_\phi$ are the canonical 
momenta conjugated to $a$ and $\phi$, respectively.

{}Following Ref.~\cite{Farajollahi:2010ni} (and also used in~\cite{Vicente:2023hba}), we introduce the 
canonical transformation
\begin{equation}
T=\frac{\phi}{P_\phi}, \qquad P_T=\frac{P_\phi^2}{2},
\end{equation}
after which the Hamiltonian becomes
\begin{equation}
\mathcal{H}_T = -\kappa^2\frac{P_a^2}{12a}+ \frac{P_T}{a^3}
+ a^3V(T,P_T).
\label{HT}
\end{equation}

As in~\cite{Vicente:2023hba}, we assume the bounce is kinetic-energy dominated\footnote{We note that by setting initial conditions far back in the contracting phase, the dominant energy density at the bounce is the kinetic density, as shown, for example in Refs.~\cite{Barboza:2020jux,Barboza:2022hng,Levy:2024naz}.}; hence, near the bounce we can neglect $V$, obtaining
\begin{subequations}
\begin{align}
a' &= -\frac{\kappa^2}{6}a^2P_a,\\
P_a' &= -\frac{\kappa^2}{12}aP_a^2+\frac{3P_T}{a},\\
P_T' &= 0,
\end{align}
\end{subequations}
where primes here denote derivatives with respect to $T$.

Imposing the Hamiltonian constraint gives the stiff-matter {}Friedmann equation
\begin{equation}
H^2=\left(\frac{a'}{a^4}\right)^2=\frac{\kappa^2}{3}\frac{P_T}{a^6},
\end{equation}
whose solution is
\begin{equation}
a(T)=a_0 e^{\pm\lambda T}, \qquad 
\lambda\equiv \kappa\sqrt{P_T/3}.
\end{equation}
This is a pure stiff-matter solution, i.e., it describes a kination regime, where the solution with the $+$ ($-$) sign is valid for $T>0$ ($T<0$).

\subsection{Quantum Model: de Broglie--Bohm Interpretation}

From the Hamiltonian~\eqref{HT}, WDW quantization proceeds by promoting 
\begin{equation}
(P_a,P_T)\rightarrow(\hat P_a,\hat P_T)=(-i\partial_a,-i\partial_T),
\end{equation}
which leads, upon imposing the Hamiltonian constraint, to the Wheeler--DeWitt (WDW) equation in the kination regime,
\begin{equation}
i\,\partial_T\Psi(a,T)=\frac{\kappa^2}{12}a^2\partial_a^2\Psi(a,T).
\end{equation}
{}From the solution $\Psi=\Omega\,e^{iS}$  given in Ref.~\cite{Vicente:2023hba}, the Bohmian trajectory follows from
\begin{equation}
P_a=\partial_a S,
\end{equation}
leading to
\begin{equation}
a(T)=a_{\rm B}\exp\!\left[\lambda T_0
\left(\sqrt{1+(T/T_0)^2}-1\right)\right],
\label{aqu}
\end{equation}
and the modified {}Friedmann equation
\begin{equation}
H^2=\frac{\kappa^2}{3}\rho\left\{1-
\frac{1}{\left[1-\frac{1}{6\lambda T_0}
\ln\left(\frac{\rho}{\rho_B}\right)\right]^2}\right\},
\end{equation}
where $T_0$ is a constant that comes from the solution of
the WDW equation and
\begin{equation}
\rho_B=\frac{3\lambda^2 m_{\rm Pl}^2}{8\pi a_{\rm B}^6}, \qquad
\rho(T)=\rho_B\left(\frac{a_B}{a(T)}\right)^6.
\label{rhoc}
\end{equation}

\section{Primordial Scalar Perturbations}
\label{section3}

We now derive the power spectrum for scalar perturbations, explicitly incorporating the modifications induced by the quantum bounce. Our approach follows a methodology analogous to that developed for LQC~\cite{Zhu:2017jew}, but critically adapted to the distinct dynamics of the dBB bounce and in the canonical variables defined in the previous section.
{
We assume that any pre-stiff-matter contracting phase occurs sufficiently far from the bounce so that its effects on observable modes are exponentially diluted. This behavior is generic in long-lived nonsingular contracting cosmologies, where modes spend extended periods outside the horizon and lose sensitivity to earlier phases~\cite{Peter:2008qz,Finelli:2001sr}.
}

{}For scalar perturbations, the {}Fourier modes $\mu_k(\eta)$ obey the Mukhanov-Sasaki equation~\cite{Mukhanov:2005sc},
\begin{eqnarray}\label{modes}
\mu_k''(\eta) + \left[k^2 - \frac{a''(\eta)}{a(\eta)}\right] \mu_k(\eta) = 0,
\end{eqnarray}
where $\eta$ is the conformal time, $\mu_k = z \mathcal{R}$, with $z = a \dot{\phi}/H$, and $\mathcal{R}$ is the comoving curvature perturbation.

The primary challenge is to evolve quantum vacuum fluctuations through the non-singular bounce and into the subsequent inflationary epoch. The pre-inflationary dynamics near the bounce imprint a characteristic signature on the perturbation modes that persists throughout the subsequent evolution. To model this process, we segment the history into three distinct dynamical regimes where the background evolution can be well-approximated, thus permitting a tractable analytical treatment for the perturbations:

\begin{itemize}
\item \textbf{A. Bounce Phase:} Governed by quantum gravitational effects (e.g., stiff-matter domination in our model), where the initial conditions for the perturbations are set.
\item \textbf{B. Transition Phase:} The interval between the bounce and the onset of slow-roll inflation, where neither the bounce-era kinetic energy nor the inflationary potential energy is fully dominant.
\item \textbf{C. Inflationary Phase:} The standard slow-roll regime, during which perturbations are amplified and exit the Hubble horizon.
\end{itemize}

The equation of motion for the modes cannot be solved analytically across the entire evolution. We therefore solve Eq.~\eqref{modes} separately in each of the three phases using appropriate approximations for the dynamics. The complete solution is constructed by matching both the mode functions $\mu_k(\eta)$ and their first derivatives $\mu_k'(\eta)$ at the boundaries between consecutive phases.

In the following subsections, we detail the solutions for each phase, implement this matching procedure, and finally compute the primordial power spectrum $\mathcal{P}_\mathcal{R}(k)$, which encodes the combined imprints of both the quantum bounce and the subsequent inflationary phase.

\subsection{Bounce phase}

The scale factor for our model is given by Eq.~\eqref{aqu}, which is expressed as a function of $T$.
However, the explicit expression for $a''(\eta)/a(\eta)$ in Eq.~\eqref{modes} is given in terms of the conformal time $\eta$. To relate $a(\eta)$ with $a(T)$, we can start by using that
\begin{eqnarray}
\label{dTdt}
\dot T = \{T,\mathcal{H}_T\} = \frac{1}{a^3},
\end{eqnarray}
where $\mathcal{H}_T$ is given by Eq.~\eqref{HT} and from which we then obtain $dt=a^3dT$. Using $dt=a\, d\eta$, we find that
\begin{eqnarray}
\label{teta}
d\eta=a^2dT,
\end{eqnarray}
which by integration gives
\begin{eqnarray}
\eta(T)
-
\eta_{\rm B}
=
\int\limits_0^T a^2(u)du
,\qquad
\eta_{\rm B}=\eta(0).
\end{eqnarray}
However, it is still not trivial to analytically obtain $a(\eta)$ directly from $a(T)$. We do this procedure through numerical methods. 
To compute the term $a''(\eta)/a(\eta)$, present in the perturbation equation, Eq.~\eqref{modes},
we first make use of Eq.~\eqref{teta}, which allows
one to express $a''(\eta)/a(\eta)$ in terms of the variable $T$. Explicitly, we find
\begin{eqnarray}
{\cal V}(T)
&\equiv&
\left(\frac{a''}{a}\right)(T)
\nonumber \\
&=&
\frac{\lambda e^{-4\lambda T_0 \left(\sqrt{1 + \frac{T^2}{T_0^2}} - 1\right)}
}{a_{\rm B}^4 T_0^3}
\frac{
\left(1-\lambda T_0 \sqrt{1 + \frac{T^2}{T_0^2}}\frac{T^2}{T_0^2}  \right)
} 
{\left(1 +\frac{T^2}{T_0^2}\right)^{3/2}}. 
\nonumber \\
\label{VT}
\end{eqnarray}
The expression (\ref{VT}) is still too complicated to allow the equation for the modes \eqref{modes} to be analytically solved. To avoid this difficulty, we first observe that ${\cal V}(T)$ can be well approximated as a P\"oschl-Teller function given by 
\begin{eqnarray}\label{VPT}
{\cal V}_{\rm PT}(\eta) 
= \mathcal{V}_0 \sech^2[\alpha (\eta-\eta_{\rm B})],
\end{eqnarray}
where $\alpha$ and $\mathcal{V}_0$ are free parameters fixed in such a way that Eq.~\eqref{VPT} can become a good approximation to Eq.~\eqref{VT}. Since we can analytically solve the modes equation Eq.~(\ref{modes}) for a function like (\ref{VPT}), we expect this to provide a good approximation for the solution in the bounce phase.

\begin{figure}[htb!]
\centering
\includegraphics[height=5cm,keepaspectratio]{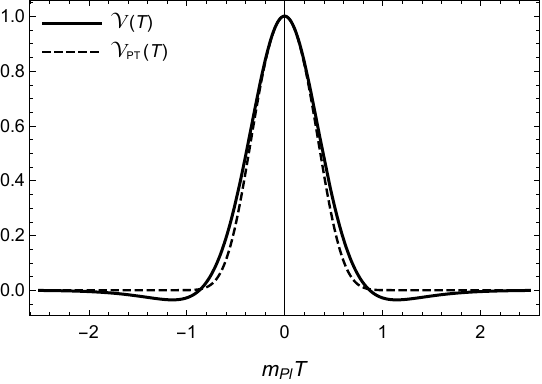}
\caption{Comparison between the exact function ${\cal V}(T)$, Eq.~(\ref{VT}), with the P\"oschl-Teller approximation ${\cal V}_{\rm PT}(T)$
for the representative parameters $\lambda=m_{\rm Pl}$,
$T_0=m_{\rm Pl}^{-1}$
and
$a_{\rm B}=1$.}
\label{fig:PT}
\end{figure}

To obtain the parameters in Eq.~(\ref{VPT}), we set the height $\mathcal{V}_0$ and the curvature $-2\alpha^2\mathcal{V}_0$ of ${\cal V}_{\rm PT}(\eta)$ at the bounce ($\eta=\eta_{\rm B}$) to those of ${\cal V}(T)$ at the bounce ($T=0$). 
Through this matching, the results for $\alpha$ and $\mathcal{V}_0$ can then be found and are given by
\begin{eqnarray}\label{alphaV0}
{\cal V}_0=\frac{\lambda}{a_{\rm B}^4T_0}
,
\quad\alpha
=
\sqrt{\frac{3}{2}}
\frac{\sqrt{1+2\lambda T_0}}{a_{\rm B}^2T_0}.
\end{eqnarray}
In {}Fig.~\ref{fig:PT} we compare the numerical result for ${\cal V}(T)$, obtained from Eq.~(\ref{VT}),  with the approximation ${\cal V}_{\rm PT}(\eta)$, Eq.~\eqref{VPT}, where in the latter we consider $\eta=\eta(T)$, Eq.~\eqref{teta}, thus expressing both functions in terms of the time variable $T$. The results in {}Fig.~\ref{fig:PT} show that Eq.~\eqref{VPT} is indeed a good approximation to Eq.~(\ref{VT}). 

Consequently, the solution for Eq.~\eqref{modes} using the approximation ${\cal V}_{PT}(\eta)$ for ${\cal V}(T)$ and with the parameters $\alpha$ and  ${\cal V}_0$ given by Eq.~\eqref{alphaV0}, is found to be given by
\begin{eqnarray}\label{modes_bounce}
\mu_k^{(\mathrm{PT})}(\eta) 
&=& 
c_k \, x(\eta)^{i k/(2\alpha)} 
\left[1 - x(\eta)\right]^{-i k/(2\alpha)} \nonumber\\
&\times& 
{}_2F_1\left[a_1 - a_3 + 1,\, a_2 - a_3 + 1,\, 2 - a_3,\, x(\eta)\right] \nonumber\\
&+& 
d_k 
\left\{x(\eta)\left[1-x(\eta)\right]\right\}^{-i k/(2\alpha)} \nonumber\\
&\times& 
{}_2F_1\left[a_1,\, a_2,\, a_3,\, x(\eta)\right],\nonumber\\
\end{eqnarray}
where ${}_2F_1$ is a hypergeometric function, $c_k$ and $d_k$ are arbitrary constants and
\begin{subequations}
\label{eq:params}
\begin{eqnarray}
a_1 
&=& 
\frac{1}{2} 
- 
i\sqrt{\frac{2}{3}}
\frac{\,a_{\rm B}^2\,T_0}{\sqrt{1+2\lambda T_0}}k
+ 
\frac{1}{2} \sqrt{\frac{3-2\lambda T_0}{3+6\lambda T_0}}
, \,\,\,\,\,\,\,\,\,\,\,\,\\
a_2 &=& \frac{1}{2} 
- 
i\sqrt{\frac{2}{3}}
\frac{\,a_{\rm B}^2\,T_0}{\sqrt{1+2\lambda T_0}}k
-
\frac{1}{2} \sqrt{\frac{3-2\lambda T_0}{3+6\lambda T_0}}, \\
a_3 &=& 
1 
-
i\sqrt{\frac{2}{3}}
\frac{\,a_{\rm B}^2\,T_0}{\sqrt{1+2\lambda T_0}}k, \\
x(\eta) &=& \frac{1}{e^{-\left[\sqrt{6}
\sqrt{1+2\lambda T_0}/(a_{\rm B}^2T_0)\right]\eta} + 1}. 
\end{eqnarray}
\end{subequations}

\subsection{Transition phase}

During the transition phase, the Universe smoothly transits from an equation of state with $\omega=1$ (kination) to $\omega=-1$ (inflation). 
Hence, we can consider $k^2\gg a''(\eta)/a(\eta)$ in Eq.~\eqref{modes} and the solution for the modes 
$\mu_k$ is simply plane-wave like,
\begin{eqnarray}\label{modes_transition}
\mu_k(\eta)
=
\frac{\tilde{a}_k}{\sqrt{2k}}
e^{-ik\eta}
+
\frac{\tilde{b}_k}{\sqrt{2k}}
e^{ik\eta},
\end{eqnarray}
where $\tilde{a}_k$ and $\tilde{b}_k$ are arbitrary coefficients.

\subsection{Inflationary phase}

After the transition phase ceases, the Universe enters the slow-roll inflationary phase. The equation of motion Eq.~\eqref{modes} for the modes in this phase becomes~\cite{Zhu:2017jew}
\begin{eqnarray}\label{modes_SRA}
\mu_k''(\eta) 
+
\left(
k^2 
- 
\frac{z''}{z}
\right) 
\mu_k(\eta) 
= 0,    
\end{eqnarray}
where $z''/z=(\nu^2(\eta)-1/4)/\eta^2$. 
The parameter $\nu$ is approximately given by $\nu\approx3/2 + 3\,\epsilon_V- \eta_V$,  where $\epsilon_V$ and $\eta_V$ are the usual slow-roll parameters in inflation, which can be approximately treated as constants during the slow-roll inflation regime.
Therefore, we can approximate $z''/z\propto 1/\eta^2$ and the equation for the modes can be solved analytically. The solution is
\begin{eqnarray}\label{modes_slowroll}
\!\!\!\!\!\!\!\!\!\!\!\!\!\!\!
\mu_k(\eta)
\simeq
\frac{\sqrt{-\pi\eta}}{2}
\left[
\alpha_k H^{\rm (1)}_{\nu}(-k\eta)
-
\beta_k  H^{\rm (2)}_{\nu}(-k\eta)
\right],
\end{eqnarray}
where 
$\alpha_k$ and $\beta_k$ are arbitrary constants,
$H^{\rm (1)}_{\nu}(z)$
and
$H^{\rm (2)}_{\nu}(z)$
are the Hankel functions of the first and second kinds, respectively.

\subsection{Matching the solutions}

Let us now determine all the arbitrary coefficients that appear in the solutions obtained above in each of the three phases. We are in particular interested in the coefficients $\alpha_k$ and $\beta_k$ of the slow-roll inflationary modes, Eq.~\eqref{modes_slowroll},
resulting from the whole evolution beginning far before the bounce, in the contracting phase.
All the coefficients are obtained by matching the solutions
and are explained below.

When starting the evolution far before the bounce, no quantum effects are present and we can impose the Bunch-Davies vacuum solution~\cite{Baumann:2009ds}
\begin{eqnarray}\label{modes_BD}
\mu_{k}^{\rm (BD)}=\frac{e^{-ik\eta}}{\sqrt{2k}}.    
\end{eqnarray}
To begin the matching procedure, we need to calculate the bounce modes, Eq.~\eqref{modes_bounce}, in the early contracting phase, which means $\eta-\eta_{\rm B}\ll 0$. In this limit, we find that
\begin{subequations}
\begin{eqnarray}
&&\!\!\!\!\!\!\!\!\!\!\!\!\!\!\!\!\!\!\!\!\!\!
x
\sim 
e^{
\frac{\sqrt{6}\sqrt{1+2\lambda T_0}}
{a_{\rm B}^2T_0}
(\eta-\eta_{\rm B})
}\to 0,\\
&&\!\!\!\!\!\!\!\!\!\!\!\!\!\!\!\!\!\!\!\!\!\!
x^{ik \frac{a_{\rm B}^2T_0}{\sqrt{6}\sqrt{1+2\lambda T_0}}
}(1-x)^{-ik \frac{a_{\rm B}^2T_0}{\sqrt{6}\sqrt{1+2\lambda T_0}}}
\sim e^{ik(\eta-\eta_{\rm B})},\\
&&\!\!\!\!\!\!\!\!\!\!\!\!\!\!\!\!\!\!\!\!\!\!
[x(1-x)]^{-ik \frac{a_{\rm B}^2T_0}{\sqrt{6}\sqrt{1+2\lambda T_0}}}\sim e^{-ik(\eta-\eta_{\rm B})}.    
\end{eqnarray}
\end{subequations}
{}From these limits and from the fact that ${}_2F_1(a_1,\, a_2,\, a_3,\, 0)=1$, the bounce mode equation given by Eq.~\eqref{modes_bounce} becomes
\begin{eqnarray}\label{modes_bounce_early}
\!\!\!\!\!\!\!\!\!\!\!\!\!\!\!\!
\lim_{\eta-\eta_{\rm B}\to-\infty}\mu_k^{(\mathrm{PT})}(\eta) 
&=& 
c_k \, e^{ik(\eta-\eta_{\rm B})} 
+ 
d_k \,
e^{-ik(\eta-\eta_{\rm B})}.
\end{eqnarray}
The latter equation is the solution for the modes in the early contracting phase, which must coincide with Eq.~\eqref{modes_BD}. Comparing both equations, one obtains
that the coefficients $c_k$ and $d_k$ are
\begin{eqnarray}\label{coeffs_bounce}
c_k=0\quad, \quad d_k=\frac{e^{ik\eta_{\rm B}}}{\sqrt{2k}}.    
\end{eqnarray}

We now match the bounce modes solution in the far past with the one in the next phase,  the transition phase. This is done by computing the bounce modes, Eq.~\eqref{modes_bounce}, for $\eta-\eta_{\rm B}\gg 0$. To do this, we first note that in the limit $\eta-\eta_{\rm B}\gg 0$,
\begin{subequations}
\begin{eqnarray}
&&
x
\sim 
1-e^{
\frac{-\sqrt{6}\sqrt{1+2\lambda T_0}}
{a_{\rm B}^2T_0}
(\eta-\eta_{\rm B})
}\to 1,\\
&&
1-x
\sim e^{
\frac{-\sqrt{6}\sqrt{1+2\lambda T_0}}
{a_{\rm B}^2T_0}
(\eta-\eta_{\rm B})
}.    
\end{eqnarray}
\end{subequations}
Hence, the bounce modes result Eq.~\eqref{modes_bounce} for $\eta-\eta_{\rm B}\gg 0$ can be simplified  
and we find
\begin{widetext}
\begin{eqnarray}\label{modes_bounce_late}
\!\!\!\!\!\!\!\!\!\!\!\!\!\!\!\!\!\!\!\!\!
\lim_{\eta-\eta_{\rm B}\to\infty}\mu_k^{(\mathrm{PT})}(\eta) 
&=& 
\left[
c_k
\frac
{\Gamma(2-a_3)\Gamma(a_1+a_2-a_3)}
{\Gamma(a_1-a_3+1)\Gamma(a_2-a_3+1)}
+d_k
\frac
{\Gamma(a_3)\Gamma(a_1+a_2-a_3)}
{\Gamma(a_1)\Gamma(a_2)}
\right]
\, e^{-ik(\eta-\eta_{\rm B})} \nonumber\\
&+& 
\left[
c_k
\frac
{\Gamma(2-a_3)\Gamma(a_3-a_1-a_2)}
{\Gamma(1-a_1)\Gamma(1-a_2)}
+d_k
\frac
{\Gamma(a_3)\Gamma(a_3-a_1-a_2)}
{\Gamma(a_3-a_1)\Gamma(a_3-a_2)}
\right]
\,
e^{ik(\eta-\eta_{\rm B})}.
\end{eqnarray}
\end{widetext}

We now match the solution (\ref{modes_bounce_late}) with the transition phase solution, Eq.~\eqref{modes_transition}. Using the coefficients $c_k$ and $d_k$ given by Eq.~\eqref{coeffs_bounce}, we obtain the coefficients appearing in the transition phase solution Eq.~(\ref{modes_transition}),
\begin{subequations}
\begin{eqnarray}
&&\tilde{a}_k
=
\frac
{\Gamma(a_3)\Gamma(a_1+a_2-a_3)}
{\Gamma(a_1)\Gamma(a_2)}
e^{2ik\eta_{\rm B}}
\\ 
&&\tilde{b}_k
=
\frac
{\Gamma(a_3)\Gamma(a_3-a_1-a_2)}
{\Gamma(a_3-a_1)\Gamma(a_3-a_2)}.
\end{eqnarray}
\label{coeffs_transition}
\end{subequations}

{}Finally, we must match the transition phase solution with the inflationary one. {}For this, 
we can start by expanding the solution of Eq.~\eqref{modes_slowroll} 
for $-k\eta\to\infty$. Using the asymptotic form for the Hankel functions~\cite{NIST}, 
we find that Eq.~\eqref{modes_slowroll} becomes
\begin{eqnarray}\label{modes_slowroll_early}
\lim_{-k\eta\to\infty}
\mu_k(\eta)
&=&
\frac{\alpha_k}{\sqrt{2k}}
e^{-i\left(1+2\nu\right)\pi/4}
e^{-ik\eta}
\nonumber \\
&+&  
\frac{\beta_k}{\sqrt{2k}}
e^{i\left(1+2\nu\right)\pi/4}
e^{ik\eta}
\end{eqnarray}
Matching the latter solution with the transition phase solution, Eq.~\eqref{modes_transition}, whose coefficients are given by Eqs.~\eqref{coeffs_transition}, one obtains
\begin{subequations}\label{coeffs_inflation}
\begin{eqnarray}
\alpha_k
&=&
\frac
{\Gamma(a_3)\Gamma(a_1+a_2-a_3)}
{\Gamma(a_1)\Gamma(a_2)}
e^{2ik\eta_{\rm B}}
\label{alphaksol}
,
\,\,\,\,\,\,\,\,\,\,
\\
\beta_k
&=&
\frac
{\Gamma(a_3)\Gamma(a_3-a_1-a_2)}
{\Gamma(a_3-a_1)\Gamma(a_3-a_2)}
e^{i\pi}
.
\end{eqnarray}
\end{subequations}
In the calculation for the power spectrum, which is to be performed next, the phase factors $e^{\pm i(1+2\nu)\pi/4}$ in Eq.~\eqref{modes_slowroll_early} does not contribute and can be omitted. 
Additionally, the results are essentially like the ones obtained in~\cite{Zhu:2017jew}, but with different coefficients, which are given here by the ones shown in Eq.~\eqref{eq:params}. 

\subsection{Power Spectrum}

The scalar power spectrum is defined by
\begin{eqnarray}\label{power_spectrum_definition}
\mathcal{P}_{\mathcal{R}}(k)
\equiv
\frac{k^3}{2\pi^2}\left|\mathcal{R}_k(\eta)\right|^2
=
\frac{k^3}{2\pi^2}
\left|\frac{\mu_k(\eta)}{z(\eta)}\right|^2.
\end{eqnarray}
The relevant modes are those in the slow-roll inflationary phase, Eq.~\eqref{modes_slowroll}, in which case we can apply the limit $-k\eta \to 0^+$. Using the series approximation
for the Hankel functions~\cite{NIST}, Eq.~\eqref{modes_slowroll} becomes
\begin{eqnarray}\label{modes_slowroll_late}
\!\!\!\!\!\!\!\!\!\!\!\!\!\!\!
\mu_k(\eta)
\simeq
-i\frac{\sqrt{-\eta}}{2\sqrt{\pi}}
\left(
\alpha_k 
+
\beta_k  
\right)\Gamma(\nu)\left(\frac{-k\eta}{2}\right)^{-\nu},
\end{eqnarray}
and we find for Eq.~\eqref{power_spectrum_definition} the result
\begin{eqnarray}\label{power_spectrum_new}
\mathcal{P}_{\mathcal{R}}(k)
=
|\alpha_{{k}}+\beta_{{k}}|^{2}
\mathcal{P}_{\mathcal{R}}^{\mathrm{GR}}(k),
\end{eqnarray}
where 
\begin{eqnarray}\label{power_spectrum_GR}
\mathcal{P}_{\mathcal{R}}^{\mathrm{GR}}(k)
\equiv \frac{k^2}{4\pi^3}\left(\frac{H}{a\dot{\phi}}\right)^2\Gamma^2(\nu)\left(\frac{-k\eta}{2}\right)^{1-2\nu},
\end{eqnarray}
is the standard scalar of curvature power spectrum as derived in general relativity (GR) and  $\mathcal{P}_{\mathcal{R}}(k)$ is the power spectrum corrected by the particle/mode production term 
$|\alpha_{{k}}+\beta_{{k}}|^{2}$ as a consequence of the presence of the quantum bounce.

Using the identity $|\alpha_{{k}}|^2-|\beta_{{k}}|^{2}=1$, one obtains that
\begin{eqnarray}
\label{power_spectrum_correction}
|\alpha_k+\beta_k|^2
=
1+2|\beta_k|^2
+
2\mathrm{Re}(\alpha_k\beta_k^*),
\end{eqnarray}
and we can express Eq.~\eqref{power_spectrum_new} as
\begin{eqnarray}
    \mathcal{P}_{\mathcal{R}}(k)=(1+\Delta_k)\mathcal{P}_{\mathcal{R}}^{\mathrm{GR}}(k),
    \label{eq:PR}
\end{eqnarray}
where
\begin{eqnarray}\label{delta}
\Delta_k
=
2|\beta_k|^2
+
2\mathrm{Re}(\alpha_k\beta_k^*).
\end{eqnarray}
Using Eqs.~\eqref{alphaV0}, \eqref{coeffs_inflation} and \eqref{power_spectrum_correction}, we find that
\begin{eqnarray}
\Delta_k
&=&
\left[ 1 + \cos\left( \pi \sqrt{1 - \frac{8 c^2}{3}} \right) \right] \mathrm{csch}^{2} \left( \sqrt{\frac{2}{3}} \frac{\pi c \,k}{k_B} \right)
\nonumber\\
&-& 
\sqrt{2} \, 
\sqrt{ 
\cosh
\left( 
2\sqrt{\frac{2}{3}} \frac{\pi c \,k}{k_B} 
\right)
+ 
\cos
\left( 
 \pi \sqrt{1 - \frac{8 c^2}{3}} 
\right) 
} \nonumber\\ 
&\times&
\cos
\left( 
\frac{\pi}{2} 
\sqrt{1 - \frac{8 c^2}{3}} \right)
\cosh^{2}
\left(\sqrt{\frac{2}{3}} \frac{\pi c \,k}{k_B}  
\right)\nonumber \\
&\times&
  \cos(2 k \eta_{B} + \varphi_k),
\label{power_spectrum_correction_explicit}
\end{eqnarray}
where we have defined\footnote{
{
Notice that $k_B$ defined here sets the characteristic \emph{comoving} scale of the bounce. The corresponding
physical energy at the bounce is $E_B=k_B/a_B$. {}Fixing the normalization $a_0=1$
implies $k_B=E_B$ today; this is a convention, while the physically meaningful
quantity remains the bounce energy $E_B$.
}}
\begin{eqnarray}\label{c_def}
c=\sqrt{\frac{\lambda T_0}{1+2\lambda T_0}},   
\end{eqnarray}
\begin{eqnarray}\label{kBv2}
k_B &\equiv& \sqrt{\frac{a''(\eta_{\rm B})}{a(\eta_{\rm B})}}\equiv\sqrt{{\cal V}_0}
=\frac{1}{a_{\rm B}^2}\sqrt{\frac{\lambda}{T_0}},   
\end{eqnarray}
and
\begin{eqnarray}\label{phase_varphi_k}
\!\!\!\!\!\!\!\!\!\!\!\!\!\!\!\!
\varphi_k = 
\arctan\left\{
\frac{\Im\,[\Gamma(a_{1})\Gamma(a_{2})\Gamma^{2}(a_{3}-a_{1}-a_{2})]}
     {\Re\,[\Gamma(a_{1})\Gamma(a_{2})\Gamma^{2}(a_{3}-a_{1}-a_{2})]}
\right\}.
\end{eqnarray}

The last term in \eqref{power_spectrum_correction_explicit}, for $k \eta_B \gg 1$, oscillates very rapidly and therefore has a negligible effect when averaged over time. Thus, for any practical purpose, when computing observable quantities, the correction factor $\Delta_k$ can be well approximated by
\begin{eqnarray}\label{delta_result_c_kB}
\!\!\!\!\!\!\!\!\!\!\!\!\!\!
\Delta_k
\simeq
\left[ 1 + \cos\left( \pi \sqrt{1 - \frac{8 c^2}{3}} \right) \right] \mathrm{csch}^{2} \left( \sqrt{\frac{2}{3}} \frac{\pi c \,k}{k_B} \right).\nonumber\\
\end{eqnarray}

In the next section, we analyze the effects of the scale-dependent quantum correction factor $\Delta_k$ on the GR power spectrum arising from the presence of the quantum bounce. By examining the resulting modifications in the CMB anisotropies induced by $\mathcal{P}_{\mathcal{R}}(k)$, we can constrain the parameters $c$ and $k_B$, which are, in turn, related to the quantities emerging from the dBB interpretation quantum framework employed in this work.

\section{Analysis and Results}
\label{section4}

Before proceeding with the analysis, we briefly comment on the parameter ranges for $c$ and $k_{\mathrm{B}}$ in light of the background-level study of the same model presented in Ref.~\cite{Vicente:2023hba}. In that work, the parameter ranges were derived for $\rho_{\mathrm{B}}$, the energy density at the bounce, given in Eq.~\eqref{rhoc},  and for the parameter $T_0$. The authors assumed $a_{\mathrm{B}} = 1$ for simplicity, while here we find it more convenient to assume the scale factor today as $a_0=1$, as usual, and we keep $a_{\mathrm{B}}$ arbitrary. We recover the arbitrariness of $a_{\mathrm{B}}$ by rescaling the quantities as $\lambda \rightarrow \bar{\lambda} a_{\mathrm{B}}^3$ and $T_0 \rightarrow \bar{T}_0 / a_{\mathrm{B}}^3$. The parameter $c$ remains unchanged, taking values in the range $c \in [0.0001203,\, 0.7071070]$
(see Ref.~\cite{Vicente:2023hba}). In the following, we consider representative values of $c$ within this interval. Conversely, since $k_{\mathrm{B}}$ depends on $a_{\mathrm{B}}$, which is now arbitrary, no fixed range can be defined for $k_{\mathrm{B}}$; instead, we adopt a broad prior range for this parameter.

Our analysis incorporates the scale-dependent distortion function $\Delta_k$ into the primordial power spectrum $\mathcal{P}_{\mathcal{R}}(k)$ using a modified version of the code \texttt{CAMB}~\cite{Lewis:1999bs}. This distortion function, defined in Eq.~(\ref{delta_result_c_kB}) and discussed in the previous section, modifies the standard GR primordial spectrum according to Eq.~(\ref{eq:PR}), where we take $\mathcal{P}_{\mathcal{R}}^{\mathrm{GR}}(k) = A_s (k / k_*)^{n_s - 1}$, with the pivot scale set to $k_* = 0.05~\mathrm{Mpc}^{-1}$.

Our cosmological analysis is constrained by data from the Planck 2018 release~\cite{Planck:2019nip}. Specifically, we employ the low-$\ell$ temperature (TT) and E-mode polarization (EE) likelihoods, the high-$\ell$ temperature and polarization (TTTEEE) data derived from the \texttt{Plik} likelihood, and the CMB lensing reconstruction data.

{}For all standard $\Lambda$ cold dark matter ($\Lambda$CDM) cosmological parameters, we adopt broad linear priors. The following parameters are allowed to vary freely: the amplitude of the primordial power spectrum ($A_{\mathrm{s}}$), the scalar spectral index ($n_{\mathrm{s}}$), the physical baryon density parameter ($\Omega_{\mathrm{b}} h^2$), the physical CDM density parameter ($\Omega_{\mathrm{c}} h^2$), the angular size of the sound horizon at decoupling ($\theta_{\mathrm{MC}}$), and the reionization optical depth ($\tau$). {}For the parameter $k_{\mathrm{B}}$, we employ a logarithmic prior, given its wide dynamic range and its role in the distortion function. The specific prior ranges for $\log_{10} k_{\mathrm{B}}$ depend on the fixed value of $c$ used in the analysis:

\begin{itemize}
\item For $c = 0.003$: $\log_{10}(k_{\mathrm{B}}\, \mathrm{Mpc}) \in [-7,\, 0.3802]$
\item For $c = 0.03$: $\log_{10}(k_{\mathrm{B}}\, \mathrm{Mpc}) \in [-7,\, -0.6198]$
\item For $c = 0.3$: $\log_{10}(k_{\mathrm{B}}\, \mathrm{Mpc}) \in [-7,\, -1.6576]$
\end{itemize}
These upper limits on $\log_{10}(k_{\mathrm{B}}, \mathrm{Mpc})$ are theoretically motivated, as larger values would lead to inconsistencies with the lensing normalization within the model.

\begin{figure}[t]
\centering
\includegraphics[width=\columnwidth,keepaspectratio]{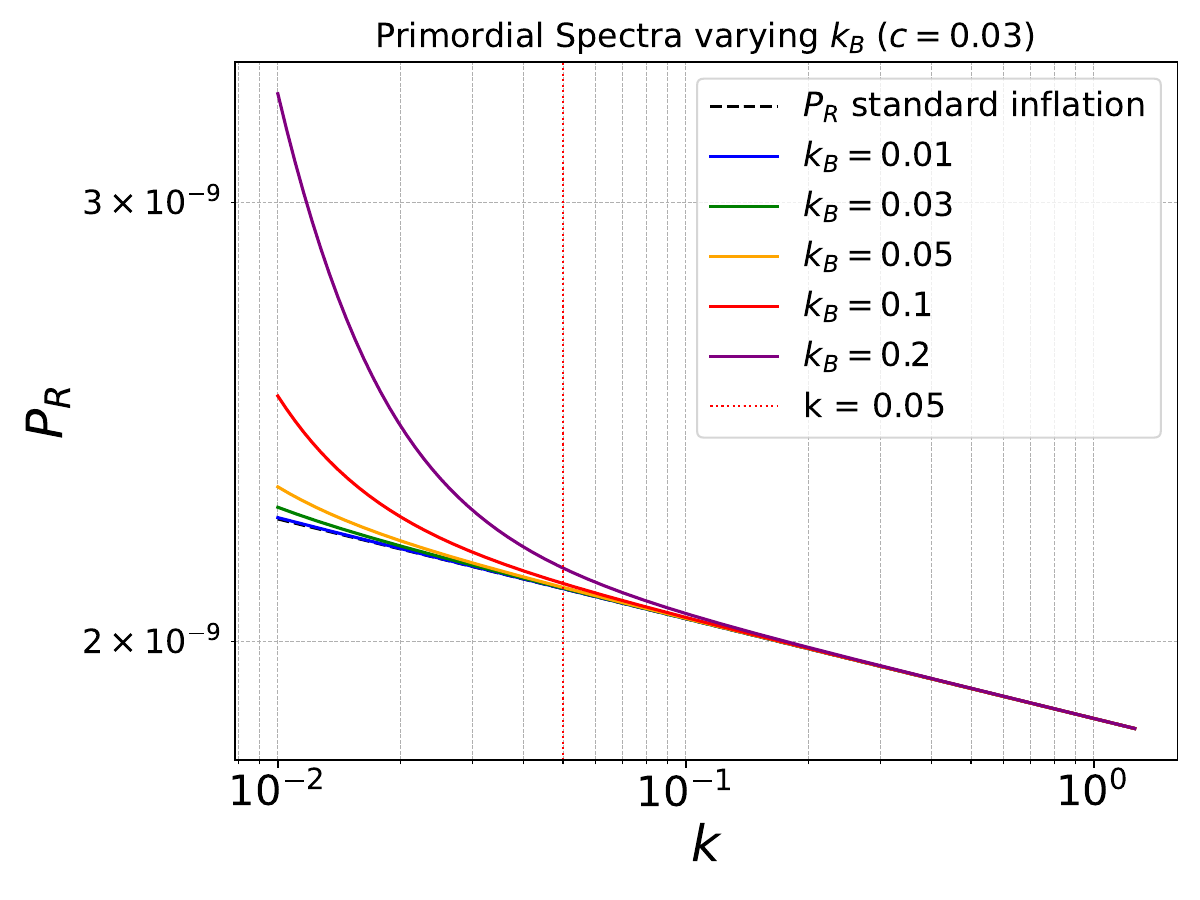} 
\caption{The primordial power spectrum as a function of the scale (in units of Mpc$^{-1}$), assuming $c=0.03$ and for different values for $k_B$.}
\label{fig:PR}
\end{figure}

\begin{figure}[t]
\centering    
\includegraphics[width=\columnwidth,keepaspectratio]{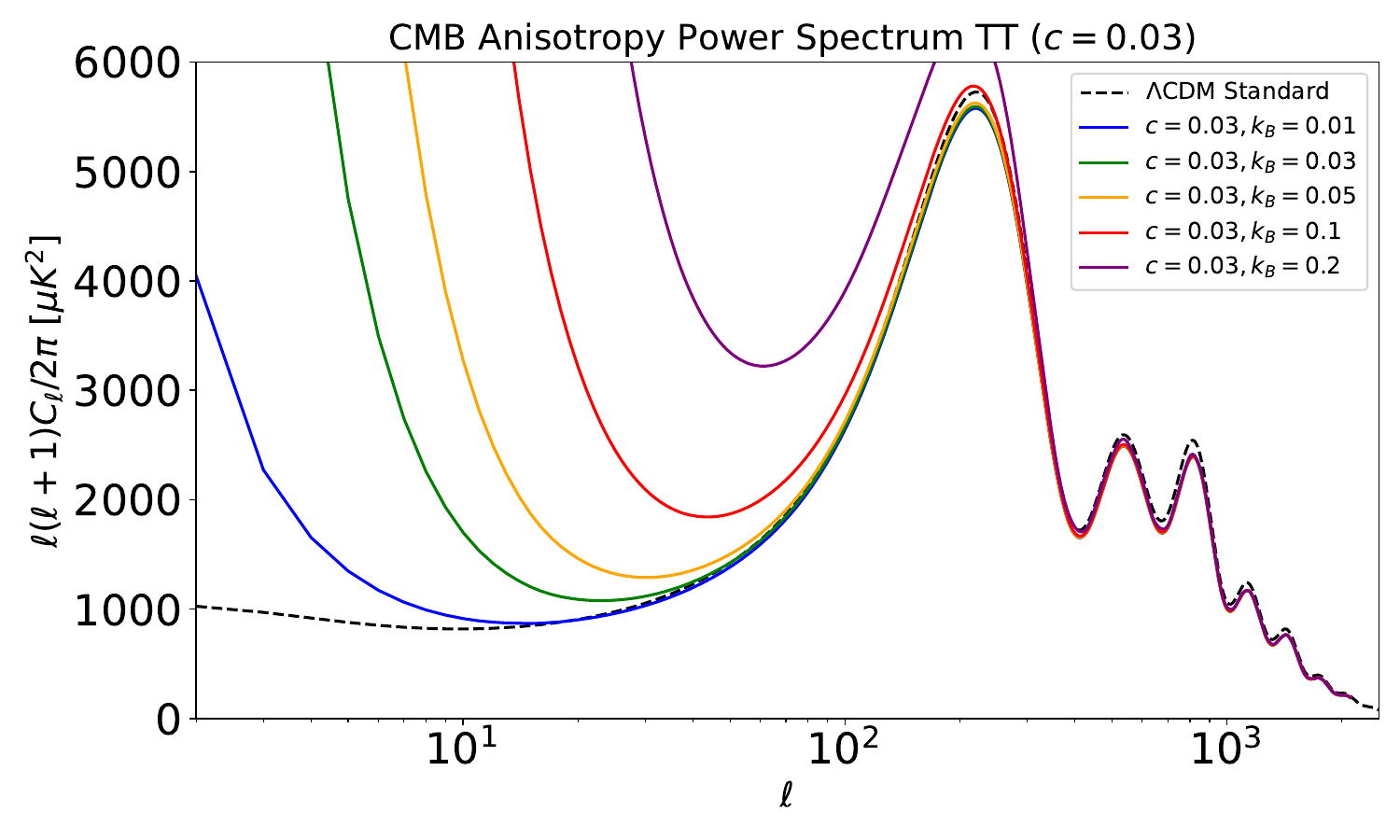}
\caption{The CMB temperature anisotropy power spectrum fixing $c=0.03$ and varying $k_B$ (in units of Mpc$^{-1}$).}
\label{fig:TT}
\end{figure}

Cosmological parameter inference is performed using the \texttt{Cobaya} MCMC sampler~\cite{Torrado:2020dgo}, which interfaces with our modified \texttt{CAMB} code to explore the parameter space. Given the strong degeneracy expected between the parameters $c$ and $k_{\mathrm{B}}$ within the distortion function, which would otherwise lead to inefficient parameter exploration, we fix $c$ to a set of predefined values of interest while varying $k_{\mathrm{B}}$. The resulting behavior of the primordial power spectrum $\mathcal{P}_{\mathcal{R}}(k)$ is shown in {}Fig.~\ref{fig:PR}, illustrating its dependence on $k_{\mathrm{B}}$ for a fixed value of $c$. {}Furthermore, {}Fig.~\ref{fig:TT} displays the corresponding CMB temperature anisotropy power spectra, obtained by incorporating our quantum cosmology (QC) modifications into the \texttt{CAMB} code.

\begin{table}[htbp] 
\caption{Mean values and $1\sigma$ errors from the analyses.}
    \label{tab:analysis_results}
\centering 
\footnotesize
\begin{tabular}{lccc}
\hline
Parameter & c=0.003 & c=0.03 & c=0.3 \\
\hline
$\log_{10}(10^{10} A_\mathrm{s})$ & $3.045 \pm 0.014$ & $3.055 \pm 0.016$ & $3.056 \pm 0.016$ \\
$n_\mathrm{s}$ & $0.9642 \pm 0.0042$ & $0.9577 \pm 0.0045$ & $0.9575 \pm 0.0043$ \\
$H_0$ & $67.31 \pm 0.54$ & $67.13 \pm 0.55$ & $67.13 \pm 0.54$ \\
$\Omega_\mathrm{m}$ & $0.3158 \pm 0.0074$ & $0.3186 \pm 0.0076$ & $0.3186 \pm 0.0075$ \\
$\sigma_8$ & $0.8118 \pm 0.0059$ & $0.8188 \pm 0.0061$ & $0.8190 \pm 0.0063$ \\
\hline
\end{tabular}
   \end{table}

\begin{figure*}[t]
\centering\includegraphics[height=15cm,keepaspectratio]{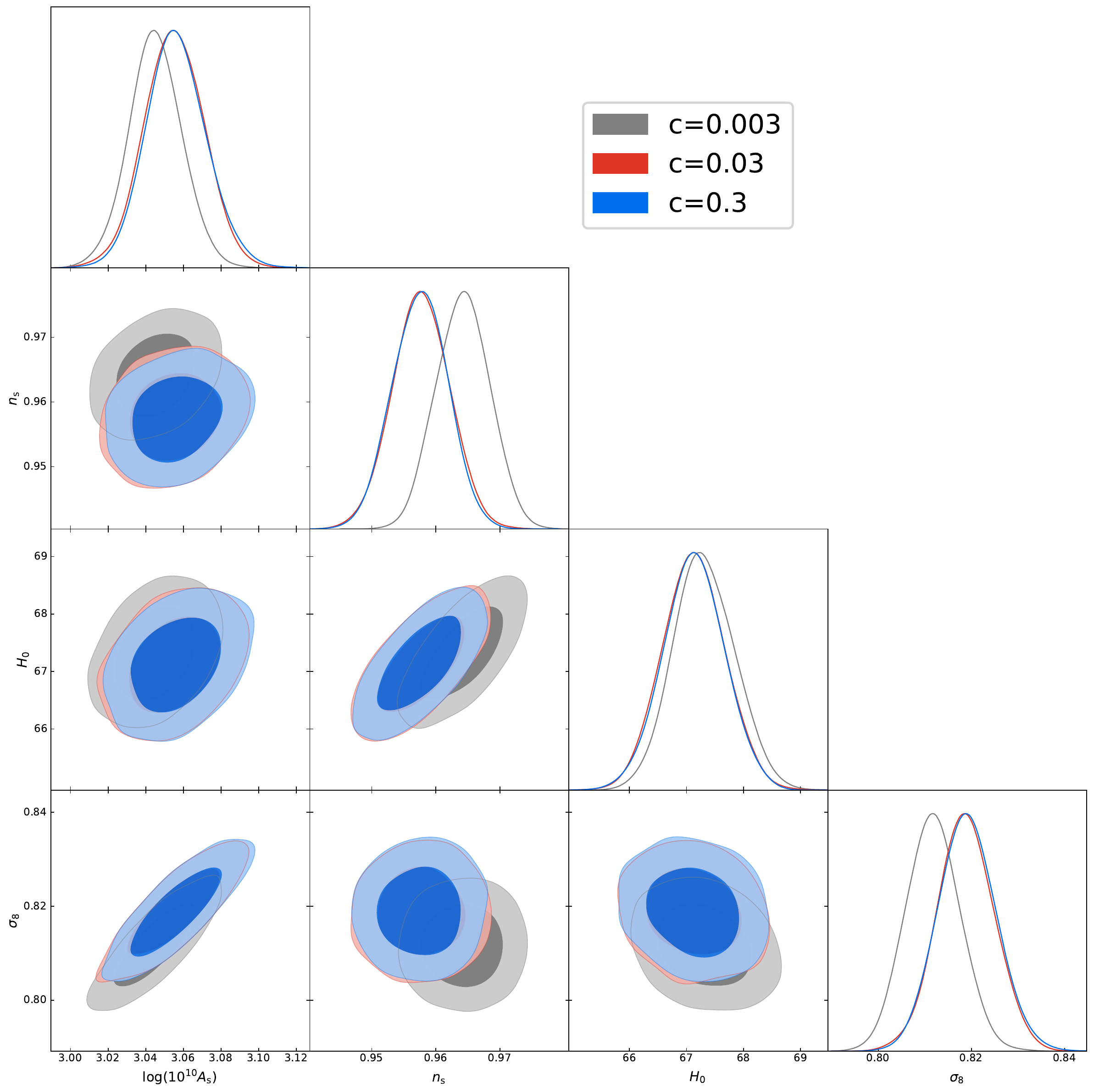} 
\caption{Confidence regions from our analysis using CMB Planck 2018 $TTTEEE$+$lowE$+lensing data \cite{Planck:2019nip}.}
\label{fig:analysis_results}
\end{figure*}

\begin{figure}[!hbt]
\centering\includegraphics[width=7.5cm,keepaspectratio]{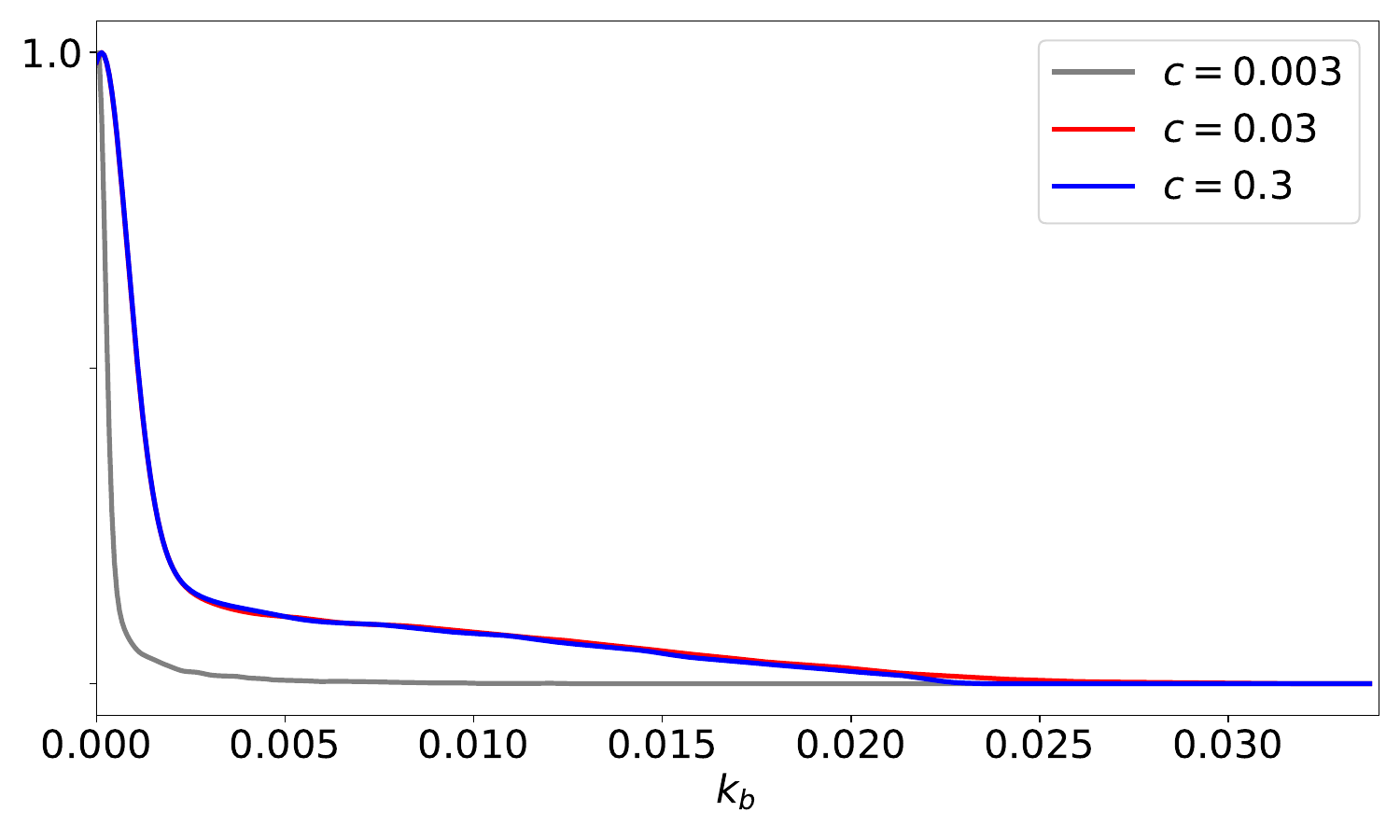}
\caption{The bounce scale $k_B$ posterior (in units of Mpc$^{-1}$) using the CMB Planck 2018 $TTTEEE$+$lowE$+lensing data \cite{Planck:2019nip}.}\label{fig:kb_posterior}
\end{figure}

The constraints obtained for the standard cosmological parameters that exhibit a Gaussian-like posterior profile are presented in Table~\ref{tab:analysis_results}, where we also report the total matter density ($\Omega_m$) and the amplitude of matter fluctuations ($\sigma_8$) parameters. The corresponding 2D confidence levels (CLs) and 1D marginalized posterior distributions are plotted in {}Fig.~\ref{fig:analysis_results}.

It can be observed that all the cosmological parameters show only minor variations, remaining consistent within $1\sigma$ with their respective values in the standard $\Lambda$CDM model \cite{Planck:2018vyg}.

The analysis indicates a slightly lower $\chi^2$ value (i.e., a marginally improved fit) when the QC model is considered. However, taking into account the inclusion of an additional parameter relative to the standard model, we conclude that, from a statistical standpoint, there is no significant evidence favoring this model. Nevertheless, the QC model remains compatible with the data, which produces, at the $2\sigma$ level, the following upper limits for $k_B$:
\begin{itemize}
    \item For $c = 0.003$: $\log_{10}(k_{\mathrm{B}}\, \mathrm{Mpc}) < -2.454$,
    \item For $c = 0.03$: $\log_{10}(k_{\mathrm{B}}\, \mathrm{Mpc}) < -1.546$,
    \item For $c = 0.3$: $\log_{10}(k_{\mathrm{B}}\, \mathrm{Mpc}) < -1.658$.
\end{itemize}
We note that the parameter $k_{\mathrm{B}}$ is rather weakly affected by variations in $c$, and that the data provide an upper limit rather than a detection. This behavior arises because smaller values of $k_{\mathrm{B}}$ recover the standard primordial power spectrum, $\mathcal{P}_{\mathcal{R}}^{\mathrm{GR}}(k)$, which remains in good agreement with the observational data. The posterior distribution of $k_{\mathrm{B}}$ is shown in {}Fig.~\ref{fig:kb_posterior}.

{
Bounce-induced features predominantly affect infrared modes with $k<k_B$, whose wavelengths today exceed the observable window. Consequently, current observations constrain only the upper bound of $k_B$, while residual bounce signatures remain observationally suppressed.
}

It should be noted that our cosmological analysis also sheds light on the well-known $H_0$-$\sigma_8$ correlation and its potential breaking within a QC model. As detailed in Table~\ref{tab:analysis_results}, for the $c=0.003$ case, our results indicate a $H_0$ value that remains consistent with those found for the $c=0.03$ and $c=0.3$ scenarios. Simultaneously, we observe a slightly lower $\sigma_8$ value compared to the higher $c$ values. This particular shift, exhibiting a tendency towards an anti-correlation between $H_0$ and $\sigma_8$, holds significant relevance in the context of current cosmological tensions~\cite{CosmoVerse:2025txj}. 

Indeed, the $H_0$ tension, characterized by a discrepancy between early-Universe (CMB-derived, e.g., $H_0 = 67.44 \pm 0.58$ km/s/Mpc from Planck 2018 TTTEEE+lensing data \cite{Planck:2018vyg}) and late-Universe (local ladder, e.g. $H_0=73.2 \pm 1.3$ km/s/Mpc \cite{Riess:2019qba}) measurements of the Hubble constant, points to the need of models favoring higher $H_0$ values. Concurrently, the $\sigma_8$ tension often refers to lower values of the matter fluctuation amplitude favored by large-scale structure (LSS) probes, particularly weak gravitational lensing measurements. This contrasts with the higher values inferred from CMB data under the standard $\Lambda$CDM model, which for Planck 2018 TTTEEE+lensing data yields $\sigma_8 = 0.8111 \pm 0.0060$ \cite{Planck:2018vyg}. {}For instance, recent weak lensing surveys like KiDS-1000 ($\sigma_8 \sim 0.740$) \cite{Heymans:2020gsg} and DES Year 3 ($\sigma_8 \sim 0.756$) \cite{DES:2021vln} consistently report $\sigma_8$ values lower than those inferred from CMB data.

Before closing this section, we note that we can use the results obtained here for $k_B$ to estimate the total number of e-folds, $N_{\mathrm{total}}$, from the bounce to today. {}From Eq.~\eqref{kBv2}, together with the scalings for $\lambda$ and $T_0$ introduced earlier, we obtain
\begin{eqnarray}
\frac{k_{\mathrm{B}}}{a_0} = e^{-N_{\mathrm{total}}} \sqrt{\frac{\bar{\lambda}}{\bar{T}_0}},
\end{eqnarray}
where $N_{\mathrm{total}} = \ln(a_0 / a_{\mathrm{B}})$. Using the ranges of $\bar{\lambda}$ and $\bar{T}_0$ (as given in Ref.~\cite{Vicente:2023hba}) and the upper limits of $k_{\mathrm{B}}$ corresponding to the chosen values of $c$, and setting $a_0 = 1$, we obtain:
\begin{itemize}
    \item For $c = 0.003$: $N_{\mathrm{total}} \gtrsim 126$,
    \item For $c = 0.03$: $N_{\mathrm{total}} \gtrsim 124$,
    \item For $c = 0.3$: $N_{\mathrm{total}} \gtrsim 124$.
\end{itemize}
Subtracting the $\sim 60$ e-folds from horizon exit during slow-roll inflation to the end of inflation, and another $\sim 60$ e-folds from the end of inflation to the present epoch, we find approximately $4$--$6$ e-folds from the bounce until the onset of slow-roll inflation. This result is consistent with Table~1 of Ref.~\cite{Vicente:2023hba}, which gives about $4$ e-folds for this period.

Note that the bound on $k_{\rm B}$ obtained from the MCMC analysis can be translated into
a restriction on the characteristic bounce energy. Using Eq.~(\ref{kBv2}), the
parameter $\lambda$ fixes the bounce energy density $\rho_{\rm B}$, which in
turn determines $k_{\rm B}$. Because $k_{\rm B}$ is observed only through an
upper bound, the corresponding implication for the number of inflationary
$e$-folds is also one-sided: only the maximum allowed value of $k_{\rm B}$
yields a lower bound on $N_{\rm total}$, as shown above. 
This also explains why the parameter ranges inferred from
Ref.~\cite{Vicente:2023hba} would not impose a stronger restriction than those
derived observationally; the observational limits on $k_{\rm B}$ are already
more stringent.

\begin{table}[htbp]
\centering
\caption{Estimated bounce energy scale $E_B$, the corresponding bounce characteristic energy $\rho_B^{1/4}$ 
and also an estimate for the expected 
lower bound on the number of $e$-folds during the inflationary phase, $N_{\rm infl}$, using the 2$\sigma$ upper limits on $\log_{10}(k_B\,{\rm Mpc})$ and representative values of $N_{\rm total}$ quoted in the text.}
\label{table2}
\vspace{0.2cm}
\begin{tabular}{c c c c}
\hline\hline
$c$ & $E_B$ (GeV) & $\rho_B^{1/4}$ (GeV) & $N_{\rm infl}$\\
\hline
0.003 & $1.18\times10^{14}$ & $1.22\times10^{15}$ & 60.6 \\
0.03  & $1.30\times10^{14}$ & $4.05\times10^{15}$ & 59.5 \\
0.3   & $1.00\times10^{14}$ & $1.18\times10^{16}$ & 59.5\\
\hline\hline
\end{tabular}
\end{table}

Adopting the results obtained above for the upper limits on $k_B$ together with the representative values 
$N_{\rm total}\gtrsim126$ for $c=0.003$ and $N_{\rm total}\gtrsim124$ for $c=0.03,0.3$, 
we show in Table~\ref{table2} the characteristic bounce energy scale $E_B=k_B/a_B$ and the maximum bounce energy 
density scale $\rho_B^{1/4}$.

Note that the results shown in Table~\ref{table2} are computed from the 2$\sigma$ upper bounds on $\log_{10}(k_B\,{\rm Mpc})$ used in the analysis (respectively $-2.454,\,-1.546,\,-1.658$ for $c=0.003,0.03,0.3$) together with $N_{\rm total}\gtrsim126$ for $c=0.003$ and $N_{\rm total}\gtrsim124$ for $c=0.03,0.3$. The estimates are order-of-magnitude upper bounds and are exponentially sensitive to $N_{\rm total}$ (changing $N_{\rm total}$ by $\pm 1$ rescales $E_B$ by $e^{\pm1}$ and $\rho_B^{1/4}$ by $e^{\pm1/2}$).
In Table~\ref{table2} we also quote an estimate for the expected lower bound on the number of $e$-folds during the inflationary phase, $N_{\rm infl}$. This estimate is 
inferred from the total number of $e$-folds from the bounce until today, $N_{\rm total}$, after subtracting both the 
expansion from the bounce to the onset of inflation and the post-inflationary expansion from the end of inflation to the present epoch. The number of $e$-folds between the bounce and the onset of inflation is generically of order $\sim 5$ 
and only weakly dependent on model details~\cite{Vicente:2023hba}. The post-inflationary 
contribution is estimated under standard assumptions for the thermal history of the Universe, including reheating and 
subsequent radiation- and matter-dominated eras\footnote{A sensible estimate is that the number of $e$-folds from the time when our present Hubble radius crossed the Hubble 
radius during inflation until the end of inflation lies in the range $[50,60]$, with the precise value depending on the 
inflationary model and reheating details~\cite{Liddle:2003as}. Under standard post-inflationary evolution with entropy conservation after
reheating, the number of $e$-folds from the end of inflation until today is also typically of order $50$--$60$.}. The resulting lower bound on $N_{\rm infl}$ is consistent with 
expectations from standard inflationary scenarios compatible with current observations.

\section{Conclusions}
\label{section5}

In this work, we have investigated the pre-inflationary Universe in the context of quantum cosmology using the de Broglie-Bohm interpretation. Our primary goal was to analyze the evolution of primordial perturbations during a quantum bounce phase and to constrain the model parameters through the observational implications of this theoretical framework.

The quantum bounce modifies the primordial scalar of curvature power spectrum by a scale dependent distortion function $\Delta_k$, Eq.~(\ref{delta_result_c_kB}).
By having this distortion function into the primordial power spectrum, we test the predictions of our model against the most recent cosmological data from the Planck 2018 mission. We found complete compatibility with the CMB data, even without a statistically significant improvement over the standard $\Lambda$CDM model. However, our study successfully constrained the parameter $k_B$, which gives the scale for the quantum bounce in our model, placing a strong upper limit on its value. 

{}Finally, we have shown a possible connection of our QC model with the $H_0$ and $\sigma_8$ tensions. By driving $H_0$ towards values more consistent with local measurements, while concurrently achieving a relatively lower $\sigma_8$ (compared to other parameter choices within our model), our QC model offers a promising avenue to reconcile these long-standing discrepancies between early- and late-Universe observations of cosmic evolution. This aligns with recent discussions in the literature, where analogous mechanisms in inflationary models have been explored to break such correlations and mitigate cosmological tensions, as discussed for example in Ref.~\cite{Rodrigues:2023kiz}.

\begin{acknowledgments}

MB acknowledge the support of {\it Istituto Nazionale di Fisica Nucleare} (INFN), {\it iniziativa specifica QGSKY}. 
We thank the use of the \texttt{CAMB} code \cite{Lewis:1999bs} and the {Cobaya} MCMC sampler \cite{Torrado:2020dgo} in this work.
R.O.R. is partially supported by research grants from Conselho
Nacional de Desenvolvimento Cient\'{\i}fico e Tecnol\'{o}gico (CNPq), Grant No. 307286/2021-5, and {}Funda\c{c}\~ao Carlos Chagas Filho de Amparo \`a Pesquisa do Estado do Rio de Janeiro (FAPERJ), Grant
No. E-26/201.150/2021. 

\end{acknowledgments}


\end{document}